# Core Logic and Algorithmic Performance Enhancements for a System Vulnerability Analysis Technique for Complex Mission Critical Systems Implementation

Matthew Tassava, Cameron Kolodjski, Jordan Milbrath, Jeremy Straub
Institute for Cyber Security Education and Research
North Dakota State University
1320 Albrecht Blvd., Room 258
Fargo, ND 58108
Phone: +1-701-231-8196
Fax: +1-701-231-8255
Email: matthew.tassava@ndsu.edu, cameron.kolodjski@ndsu.edu, jordan.milbrath@ndsu.edu
jeremy.straub@ndsu.edu

**Abstract**

Core logic and processing improvements were made to the software for operations and network attack results review (SONARR) and are presented, herein. Previous SONARR versions' Boolean-only logic, derived from the Blackboard Architecture, was replaced with generic logic that allows any .NET type (e.g., integers, decimals, strings) to be utilized within facts. This allows calculations and equality operations with all data types to drive the algorithm's processing of network models. Additionally, multi-compute capabilities were implemented to increase the processing power for larger workloads. In this paper, the new logic objects are described, examples are presented to illustrate the efficacy of creating digital-twin systems using the new generic logic, and performance test results are presented that illustrate the expanded processing capability from the multi-compute functionality.

## 1. Introduction

A system vulnerability analysis technique (SVAT) was presented in [1] that implemented an explainable rule-fact artificial intelligence (AI) as the core logic for a technology that could analyze complex mission critical systems (CMCS). The core technology for this analysis technique is the software for operations and network attack results review (SONARR).

SONARR is an SVAT tool built on the Blackboard Architecture's Boolean rule-fact logic. It takes a user-built digital twin of a real system – represented as collections of containers, links, facts, common properties, and rules – and processes it by triggering rules and changing facts along a pathway through the network model. The objects that comprise the network model represent aspects of the environment under assessment. Containers can be entities such as people, places, or things. Facts are the attributes of those entities. Links model the relationships between containers. Common properties group facts of specific types. Rules implement the performable activities in the environment.

Actively changing facts during traversal simulates the dynamic real-world environment and allows the SONARR algorithm to evaluate rules in different areas of the network that would not have been in the applicable state, if the network environment was static. The results of SONARR analysis are the paths an attacker could take through a given network, complete with information on what rules were triggered, where in the network those rules were triggered, the facts that were altered, and the order of containers the paths take. The overall objective of SONARR is to provide an explainable and automated artificial intelligence solution for system vulnerability analysis that human analysts can use to facilitate securing systems.

Over the course of development, the SONARR technology has been improved in a number of ways, such as by adding object referencing during processing [2] and implementing multi-threading [3] to expedite processing large networks. Overall, improvement of algorithm performance has been a predominant developmental focus, as shown through the enhancements and experimentation described in [3]. The upgrades introduced in this paper enhance SONARR's algorithmic performance and its core logic that is used for assessment during network processing.

## 2. Background

SONARR's logic is built upon a rule-fact logic evaluation system called the Blackboard Architecture. The Blackboard Architecture was introduced by Hayes-Roth in the 1980s [4], based on similar technology that was used in the HEARSAY-II system that was proposed in a DARPA-sponsored competition for speech recognition software [5]. The technology has been used for various applications, such as agent coordination in real-time strategy games [6], countering terrorism [7], software testing [8], medical imaging [9], [10], PROLOG programming [11], robotics development [12], [13], [14], [15], and mathematical proof creation [16].

The Blackboard Architecture uses logic similar to rule-fact expert systems AI. These systems embody the knowledge of experts, for a particular topic, as cause-effect parameters in a knowledge base. They enact changes on a fact set via rules associated with the knowledge base. Rule-fact expert systems are an inherently explainable technology that provides clear, human interpretable reasoning to reach its conclusion [17]. This makes it a promising methodology to implement eXplainble AI (XAI) that humans can trust with important tasks.

The term 'explainable', within AI, refers to human understandability of the processes used by the AI to reach its conclusion. The recent advances in machine learning have made AI a popular technology within the computing community [18]; however, many AI techniques' opaque nature makes their conclusions difficult to verify, discouraging trust in their decision-making capabilities [19], [20]. To this end, DARPA launched an initiative to develop XAI machine learning models that can provide mechanisms for humans to follow their reasoning, while still providing a "high level of learning performance" [21].

SONARR's technology implements XAI through the use and expansion of the Blackboard Architecture, an already explainable AI mechanism. It builds upon attack chain patterning, like the MITRE ATT&CK [22] and the Lockheed Martin Cyber Kill Chain [23] frameworks, to establish an XAI that can identify possible attack pathways through a given network.

## 3. Technology

The data assessed by SONARR is stored as a network made up of objects that serve various roles. Containers and links describe node-to-node relationships and provide the network's topology, while also housing facts that rules assess during processing. The facts within the containers, links, and network hold values that describe attributes and collectively represent network states. The rules utilize pre- and postconditions to assess the network facts through cause and effect-like patterns. These and other objects make up the networks that the SONARR algorithm pulls data from during assessment.

### 3.1. Network Objects

This section reviews and discusses the networks objects and their roles in the SONARR technology.

#### 3.1.1. Action

The action object allows rules to interact with the operating system running the software. This allows a rule to trigger and invoke an action that runs a command on the target operating system, performing any task achievable through the command line interface.

For example, if a user wanted the Notepad application to run any time a specific rule is triggered, an action that could perform this would have the command "notepad.exe". This is the command for PowerShell to open a notepad window. Actions run during traversal processing, when their associated rules are triggered, whether the path is a final path or not. If this action is tied to a rule that triggers four times during the SONARR traversal session, there would be four notepad windows open at the end.

Figure 1 depicts an action object with a description of "run Notepad executable" and a command string of "notepad.exe".

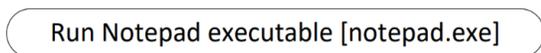

**Figure 1.** Depiction of an action object.

*3.1.2. Common Property*

The common property object is an important part of SONARR's extended rule-fact logical assessment capabilities. They allow generic rules to be truly flexible in the Blackboard Architecture by associating rule conditions with common properties that facts share, instead of having rules directly reference facts themselves. Because each fact is an object in the network, one rule can be created that applies to multiple areas of the network by applying to multiple facts.

Figure 2 depicts three string-type common properties that indicate system privilege level, the operating system, and the operating system version.

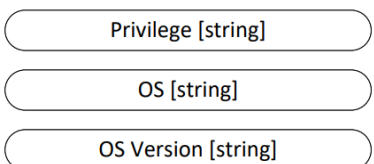

**Figure 2.** Depiction of three common property objects.

*3.1.3. Container*

The container object is an entity object that represents a node in a network. This object can represent any item in the world such as hardware, software, people, places, or groups. Attributes of the represented item are stored in the container as facts. Container objects are combined with link objects to create a network model of inter-connections and relationships that SONARR processes.

Figure 3 depicts an example of a PC container object with the name PC-1307. It has a parent container of VLAN12, and three facts describing the computer's privilege level, operating system, and operating system version.

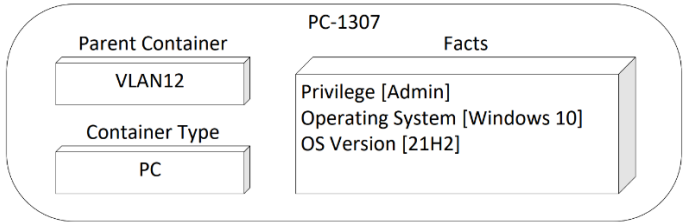

**Figure 3.** Depiction of a container object.

*3.1.4. Container Type*

The container type object associates a type and an associated image/icon with containers. The type (such as a PC, database, firewall, or person) can be used to categorize the containers of the network. These are relatively unimportant objects within the processing of a network, but they facilitate displaying visual representations of the network.

*3.1.5. Entity*

The entity object provides inheritance functionality for containers and links, both of which represent mutable elements of the network. To facilitate this, entity objects hold the facts that containers and links use as attributes during SONARR processing. Because both containers and links inherit and behave like entities within the network, they will be referred to as entities, in this paper, in some cases.

*3.1.6. Fact*

Fact objects represent attributes of an entity or environment. Facts' purpose is to store a value that is assessed, and possibly altered, through SONARR's rule-fact logic. They can be associated with a common property identifier for specific rule assessments. They are not unique to an entity, meaning that multiple entities can hold the same fact.

A fact can have any .NET value type (such as string, integer, decimal, or Boolean). A fact that is associated with a common property identifier must use the identifier's value type.

Figure 4 depicts three fact objects associated with common property identifiers. Each fact has a string value (matching the value type of their common property identifiers) that describes what the fact represents.

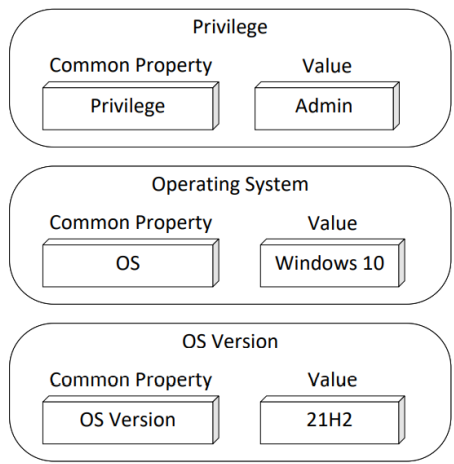

**Figure 4.** Depiction of three fact objects.

*3.1.7. Link*

Links are entity objects that represent relationships between two containers (nodes) of a network. The link object was designed as an entity, inheriting facts and properties, so that events affecting a relationship between network entities can be modeled. Additionally, the link object allows SONARR to move through the network by providing directional connections between containers.

For example, Figure 5 depicts an ethernet link between two computers, PC-1307 and PC-1309. It has a traversability value—a decimal number representing order of importance during processing—of 0.97 and facts that represent properties of the ethernet connection. The cable type fact has a string value of Cat6, the cable length fact has an integer value of 50, and the data rate fact has an integer value of 10.

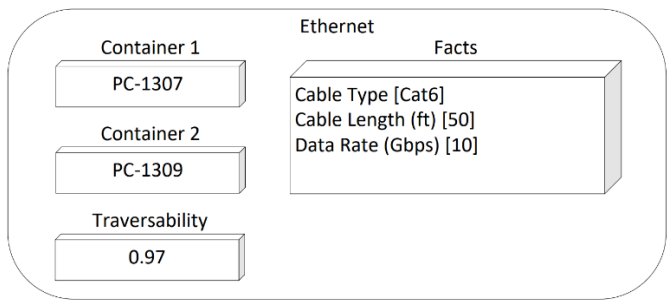

**Figure 5.** Depiction of a link object.

*3.1.8. Rule*

Rule objects provide the logic assessment within SONARR. Large collections of rules are evaluated for each container-link-container connection the SONARR algorithm processes. Evaluation of a rule begins with the preconditions, which define the fact values the network/connection must have for successful evaluation. If the preconditions return true, processing moves on to the postconditions, which enact a predefined alteration to network facts.

The confidentiality, integrity, and availability (CIA) triad is a collection of metrics used to assess the impact of vulnerabilities, exploits, and attacks [24]. These metrics are used within rules to express the impact rules can have on a target system. CIA metrics are stored inside rule objects so users can analyze the CIA impact of rules during and after traversal.

Rules' run time values express how long the activities represented by a rule would take to run in a live environment. The rule's success value is defined by the user at rule creation and indicates severity. This value is used in SONARR to order rules by severity before processing.

There are two types of rules: traversal and non-traversal rules. This is represented with a Boolean value that is set by the user at rule creation. Traversal rules are evaluated during processing and advance the algorithm across the connections that the traversal rule triggers on. Non-traversal rules solely alter the environment and do not involve movement.

Figure 6 depicts a rule that unlocks a server room door. The rule's preconditions assess the state of the link within the connection (i.e., whether it is locked and if the passcode has been obtained). The postcondition specifies what change happens if the preconditions are met (i.e., the server room door is unlocked).

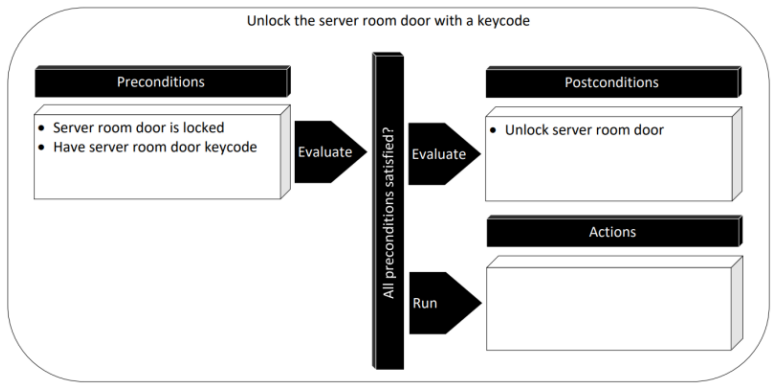

**Figure 6.** Depiction of a rule object.

### 3.1.9. Precondition

The precondition object is used, in a rule, to indicate which elements of the network/environment must have a specific value for the rule to trigger. The object is comprised of requirements and a lambda expression that is compiled into an invokable operation that evaluates to a Boolean.

The operation (compiled expression) takes an array of objects, which are collected by the SONARR algorithm based on the requirements. Genericism allows any fact type (such as integer, string, decimal, or Boolean) to be used as input for the operation. Each requirement is a tuple that specifies where (i.e., container1, link, container2, or network) the algorithm finds the fact. Those facts' values are input into the operation and evaluated. The precondition is satisfied if the operation returns true.

Figure 7 depicts a precondition that evaluates if a server door is locked. The requirements are collected from the network and input into the operation for evaluation. The precondition is satisfied due to the operation evaluating to true.

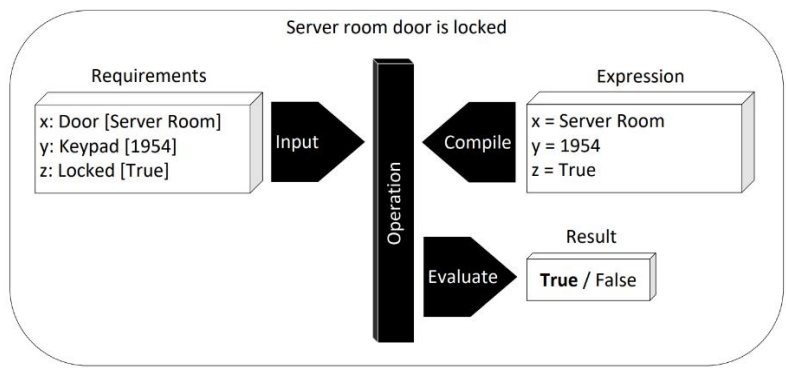

**Figure 7.** Depiction of a precondition.

### 3.1.10. Postcondition

The postcondition object is used to define an alteration that will be performed on a target fact. Just like precondition objects, postconditions utilize requirements and a compiled expression. However, instead of evaluating to a Boolean result, the postcondition's operation produces the value that the target fact will be assigned. The SONARR algorithm uses the target property of the condition to locate the target fact and sets its value to the operation's result.

Figure 8 depicts a postcondition that unlocks a server room door based on a four-digit keycode.

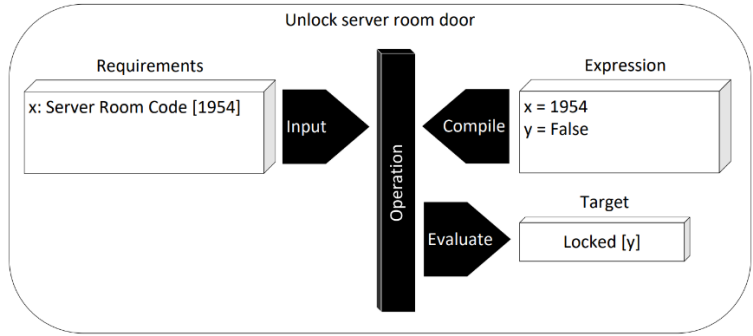

**Figure 8.** Depiction of a postcondition.

3.2. Processing Objects

This section discusses the processing objects that are instrumental to the SONARR algorithm. Their function during network assessment is also discussed.

3.2.1. Connection

The connection object is an output of the SONARR algorithm. It is used to create reality paths, which are made by chaining created connections together chronologically. It also records the network changes made via SONARR's rule-fact assessment at the container-link-container (connection) level of the network. These changes are made to the connection and appended to reality paths, so the underlying network model is not changed. Connections provide an iterative history of a reality path that can be stepped through during post-processing.

Figure 9 depicts a connection object with example variant entities, which comprise the connection itself. An example rule is also shown in the list of triggered rules.

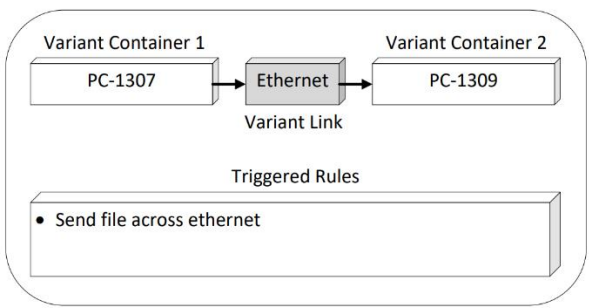

**Figure 9.** Depiction of a connection object.

3.2.2. Reality Path

The reality path object is created and output by SONARR. Each reality path represents a unique traversal that an attacker could take through a network. The reality path object is built connection-by-connection during processing. It records the rules triggered and the facts altered, as it goes. Making and recording changes within the reality path allows the original network to remain untouched. This is important because multiple reality paths are built simultaneously during traversal. Each requires a pristine base network to reference during processing.

Figure 10 depicts a reality path object with two connections. The first connection is between a room and a PC, through a keyboard, and the second connection is between a PC and another PC, via an ethernet link. This reality path describes access from development room 13 to PC-1309. The active variant facts define user privilege on PC-1307 and that trojan malware is being file shared.

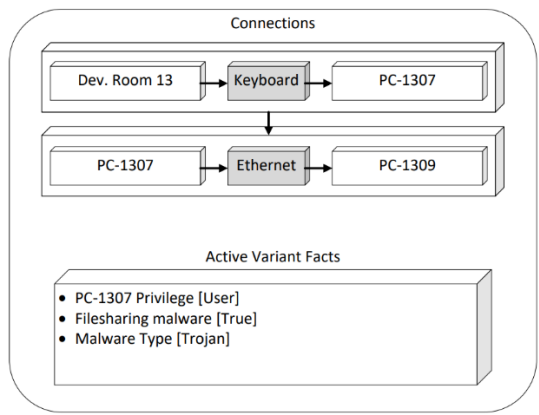

**Figure 10.** Depiction of a reality path object.

### *3.3. Algorithm*

The purpose of the SONARR algorithm is to find all routes through a network from a user-defined start container to a user-defined end container, by applying a set of rules on the network's set of facts, without cross-influencing the generated paths. There have been four algorithm versions developed for SONARR to achieve this task: the reality path string parsing algorithm [1], the in-memory single-threaded algorithm [2], the disk-writing multi-threaded algorithm [3], and the generic logic multi-compute algorithm, which is introduced in this next section.

3.3.1. Generic Logic Multi-Compute

There are two objectives for the current improvements to SONARR. The first is to implement multi-compute functionality. The second is to upgrade the rule-fact logic to allow for more than just Boolean comparisons. Thus, the SONARR technology was enhanced to perform multi-threaded processing across multiple computers with an updated generic rule-fact logic.

Multi-compute uses a relatively straightforward implementation. A system of request/response messages, over a TCP connection, was implemented to establish communication between a control node and worker nodes. The control node manages all of the connected workers and every worker request is handled by it. The worker nodes' only job is to process the work given to them and return completed paths.

The generic fact logic is more complicated because it requires fact values to be able to represent integers, decimals, strings, Boolean values, and other types. This required that the binary comparison system be transformed into an all-encompassing type comparison system. In addition to equality confirmation (==, !=), the new logic provides handling functionality for all comparison operators (<, >, <=, >=) and for performing simple calculations (+, -, *, /) on both constants and variables, before and after comparison operations. Basically, the logic needed expansion to better represent the real world.

The generic logic multi-compute algorithm used in the expansion of SONARR, which is presented in this paper, is discussed in two parts: first multi-compute is discussed in Section 3.3.1.1, then the generic logic is presented in Section 3.3.1.2.

3.3.1.1. Multi-Compute

The multi-compute algorithm uses a client/server approach where a single control node acts as a server for the worker nodes that process work through jobs utilizing multi-threaded techniques.

The worker SONARR executables are started individually through the command line and require the IP address and port of the control node they will be communicating with. The worker attempts to connect to the control node. On the control node, the user creates a scenario, enters the IP addresses and ports of the worker nodes, and starts the processing. The scenario data is sent to each worker, where the worker parses it and initializes its processing tasks.

The processing tasks are created and stored in a manager object that each node initializes. The tasks asynchronously handle assignments within the node, such as requesting additional work, handling work requests from other nodes, load balancing for the multi-threaded algorithm, and reporting/writing completed paths (dependent on node type). The manager initializes a specified number of multithread jobs (the default configuration is the number of processors in the node minus one) which run the SONARR logic on reality paths. Each job has a backlog of queued work that it processes and a collection for storing the completed paths that are found. Supplying work and handling completed paths, within each job, is done by the manager's processing task. Figures 11 and 12 depict the workflow of the processing tasks in a manager for a control node and a worker node, respectively.

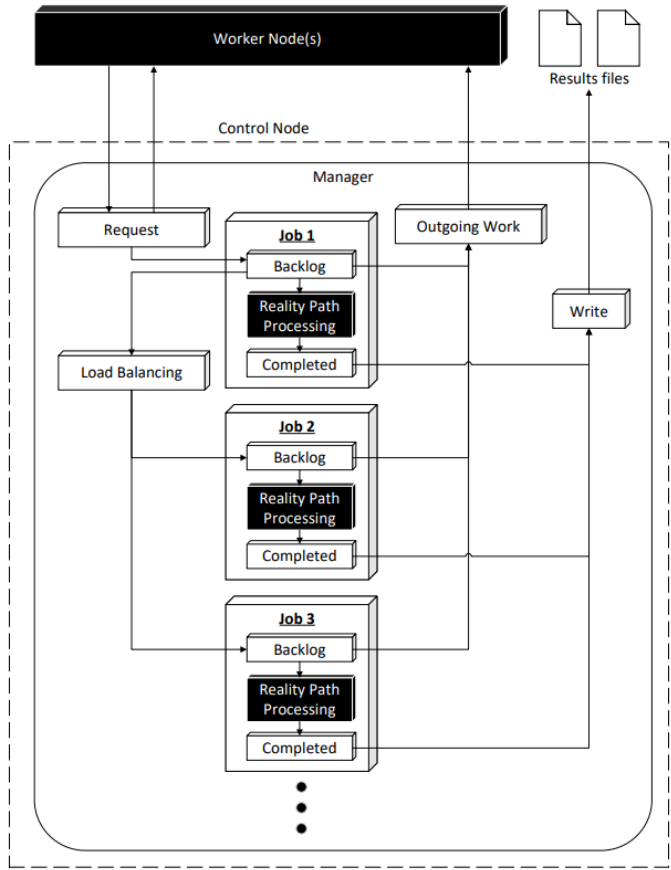

**Figure 11.** Simplified depiction of task delegation and workflow within the manager of a control node.

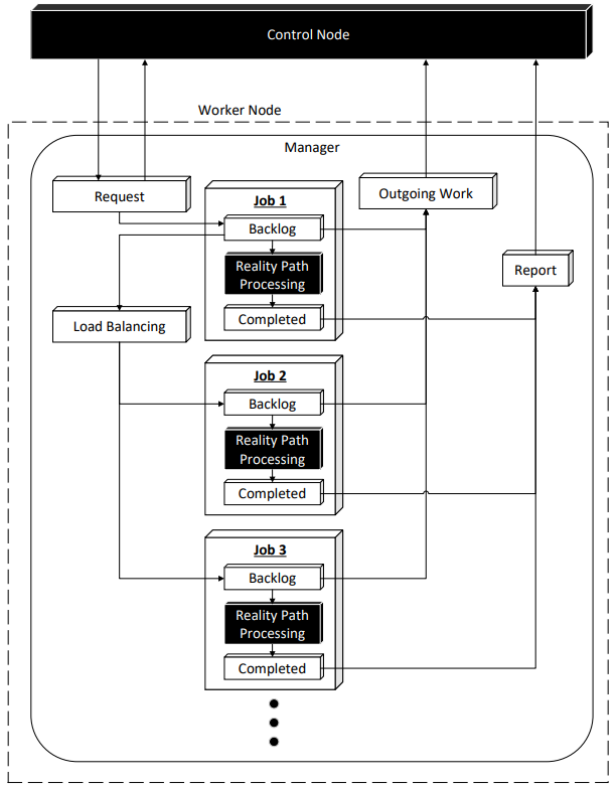

**Figure 12.** Simplified depiction of task delegation and workflow within the manager of a worker node.

The requesting task's purpose is to request work from other nodes, if its jobs are idle (i.e., they have an empty backlog and are not actively processing a path). It confirms that every job in the node is waiting for work and begins iterating through its directly connected nodes to request paths. Each neighboring node is queried until it exhausts the list or receives paths. Any received paths are added to the requesting node's first job, to be distributed by the load balancer. Figure 13 illustrates this process.

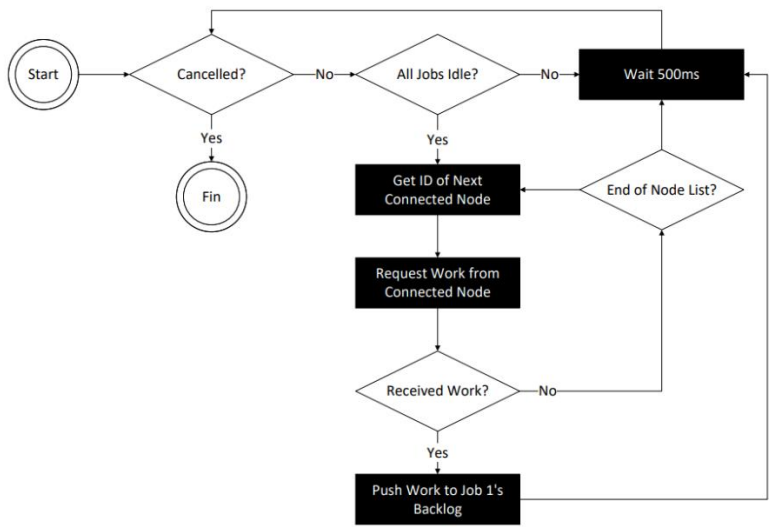

**Figure 13.** Flowchart of the request work task.

The handle requested work task (also called the outgoing work task) gets backlogged paths from its local node and sends them to another node. It does this by checking to see if there is an outgoing buffer that needs to be filled. These buffers are created when the node receives a request for more work. If buffers needing to be filled are identified, the outgoing work task attempts to fill them. This involves trying to take half of the total backlogged paths (half of the work of each job) and pushing it into the buffer. The buffer is then released and whatever it contains, empty or not, is sent to the requesting node. Figure 14 illustrates this process.

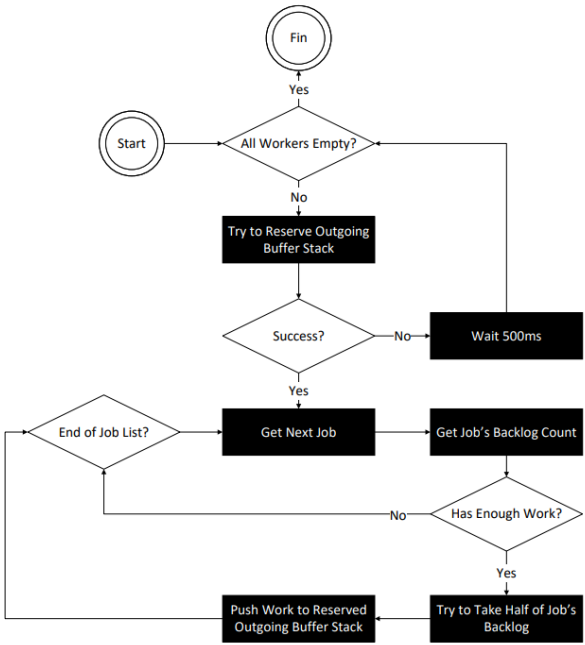

**Figure 14.** Flowchart of the outgoing work task.

The load balancing task takes half the work from the node's first job and distributes it to the rest of the jobs. Originally, the task evenly split work between all jobs that needed it, but this was too time consuming (as it required calculating which jobs were idle and how much work to give each one). Instead, using repeated attempts of taking half of the first job's backlog and giving it to another job was shown to be much faster and more reliable. Figure 15 illustrates the load balancing task's process.

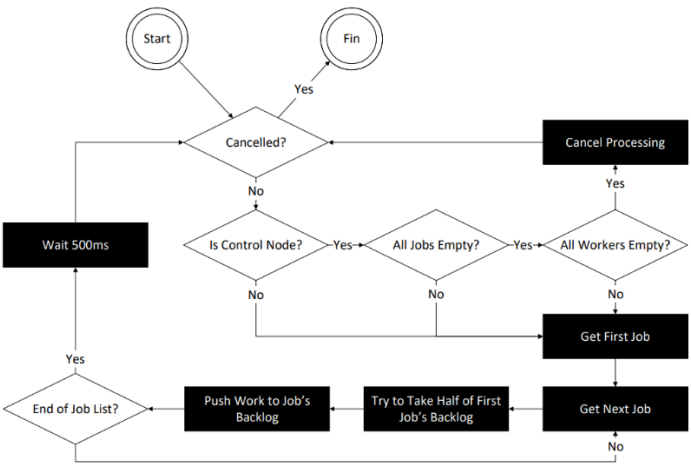

**Figure 15.** Flowchart of the load balancing task.

Another key task is the reporting task. The reporting task is only implemented within worker nodes. This task sends completed paths from the worker node to the control node. A maximum transfer size was established, so that the data sent cannot overflow buffers or the control node's general memory. Figure 16 illustrates the worker's reporting task process.

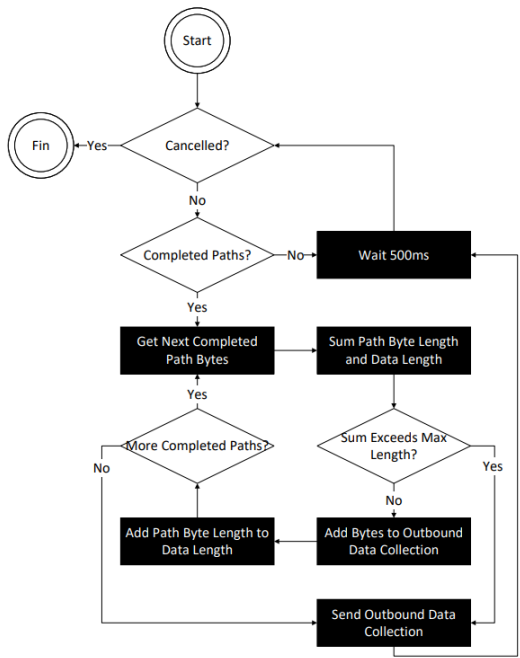

**Figure 16.** Flowchart of the worker's reporting task.

The writing task is, conversely, implemented only within control nodes. It has two components: the copying element, and the writing element. The copying element continuously pulls completed path data and attempts to push it to a large buffer. If the data it is trying to add overflows the buffer, then the buffer array's contents are copied to a temporary array and that temporary array is added to the writing collection. The large buffer is cleared, the new data is added, and the process continues. An index associated with each reality path stores the path data's starting position within the destination file for individual path retrieval. These indexes are part of the same process that the completed path data follows. However, the indexes are written to a separate file and are used to read from the path data file in sections.

The writing element is small. It constantly takes index/path data arrays from the copy element and writes them to the appropriate files. The goal of the writing task is to create large buffers so that large chunks of data can be written, with one write operation, to the disk. Experimentation showed that writing path data individually created a bottleneck waiting for the write operation. Figure 17 illustrates the writing task process.

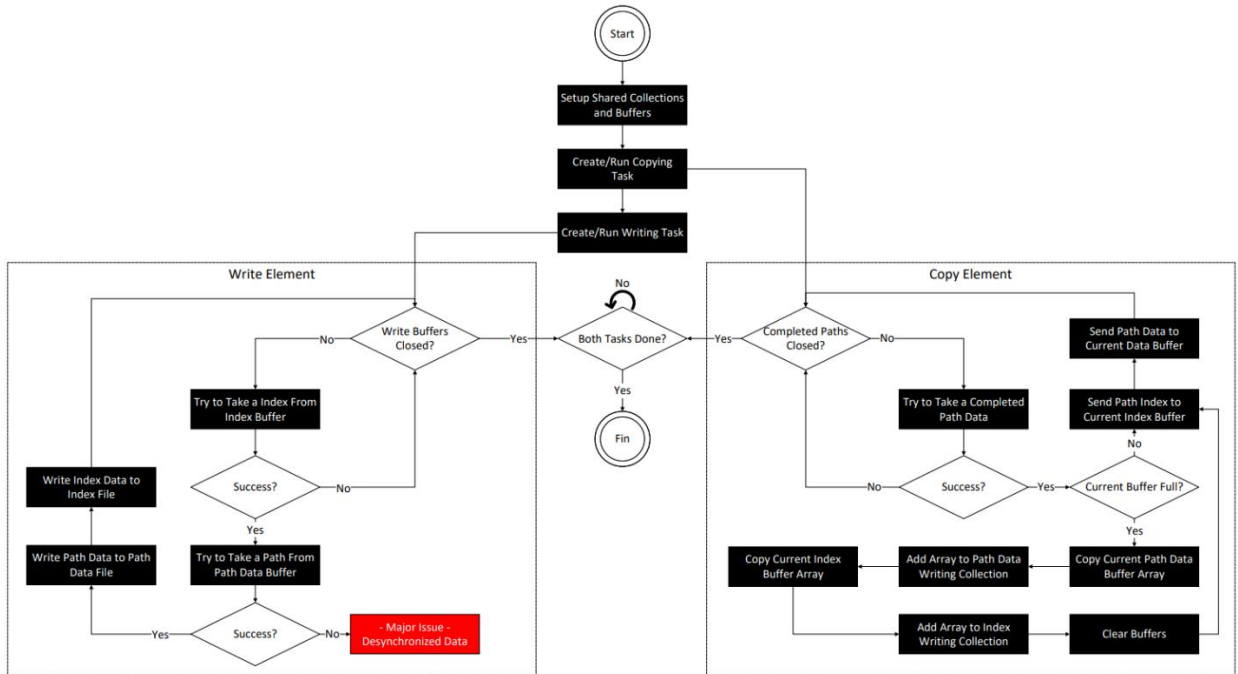

**Figure 17.** Flowchart of the control's path writing task.

The load balancing task has a check that the control node accesses. This check determines if all of the jobs within the node are empty and if all workers have an empty status. Empty worker status is recorded by the control node when a worker requests paths and the control replies with an empty array, meaning that the worker is still empty. The control node's load balancing task's check sees this, along with the empty jobs within the control itself, and trips the cancellation token. All jobs and processing tasks (except the control's writing tasks) read this cancellation token and exit their loops.

Once processing is cancelled by the control node and the loops end, a cancellation message is sent to each worker, one-by-one. This is done iteratively so that the worker nodes can gather any remaining completed paths and send them to the control. The control pushes these remaining paths to the writing task, which hasn't stopped because it's stopping criteria is not based on the cancellation token. Once all workers have confirmed their finished status, the control closes the completed paths collection that the writing task pulls from. This prevents anything else from being added to the collection and signals to the writing task that, once it is empty, it can stop. Finally, while the writing task finishes, a metric collection task is initialized to retrieve the processing data collected by each worker. One-by-one the metric data is transferred and the metric collection task agglomerates its data with the incoming data. The entire multi-compute algorithm is complete once the writing is finished for both the completed path data and the agglomerated metric data.

3.3.1.2. Generic Logic

This section discusses the changes to the logic system and how they affect the SONARR algorithm.

3.3.1.2.1. Features

The new SONARR algorithm facilitates the creation and management of complex rules. Previously, fact values could only be true or false, with rules checking and modifying the Boolean values. This Boolean-only logic limited the modeling of device properties and interactions by forcing every property or cause-effect relationship to be represented by a single or combination of several Boolean values. The new generic logic allows the representation of a device property as a Boolean, integer, decimal, or string value.

Rule pre- and postconditions can, thus, require facts to have certain Boolean, integer, decimal, or string values. Boolean and string values must be either equal or not equal to a certain value. Integer and decimal values must be equal to, not equal to, greater than, greater than or equal to, less than, or less than or equal to a certain value.

The new SONARR algorithm also handles variants differently. Facts are the only objects that can be variants, as they are used in reality paths to store the values that are produced by rule outputs (i.e., the changes). The base network is read-only, since it is shared between all paths during processing. Thus, when a fact's value is altered, the corresponding fact object in the network is not modified. Instead, the fact is cloned into the reality path's collection of active variant facts and the variant represents the change in the network. If the fact is altered again, later in processing, that alteration is applied to the active variant fact stored in the reality path. This process is shown in Figure 18.

If facts are created during scenario processing—which is explained later in this section—these new facts are added to the reality path and not the base network.

When the algorithm accesses a fact object, the reality path's active variants are checked first. If no active variant fact is found, the base network is queried, and the original read-only fact is returned. An active variant can be retrieved by either a fact ID (shown in Figure 19) or both an entity ID and a property ID (shown in Figure 20), facilitating the retrieval of condition requirements.

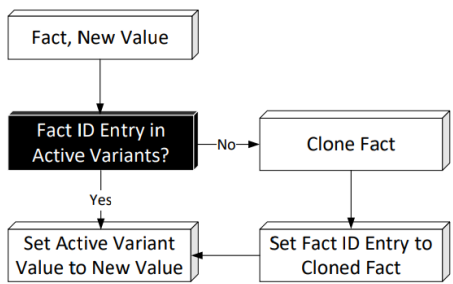

**Figure 18.** Altering a fact's value.

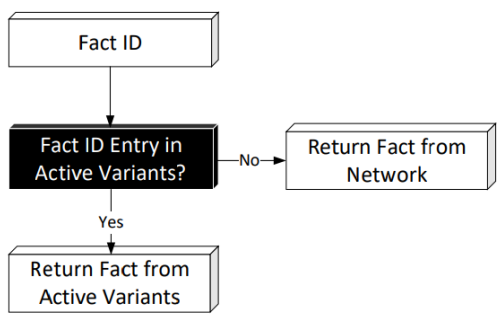

**Figure 19.** Retrieving a fact by the fact's ID.

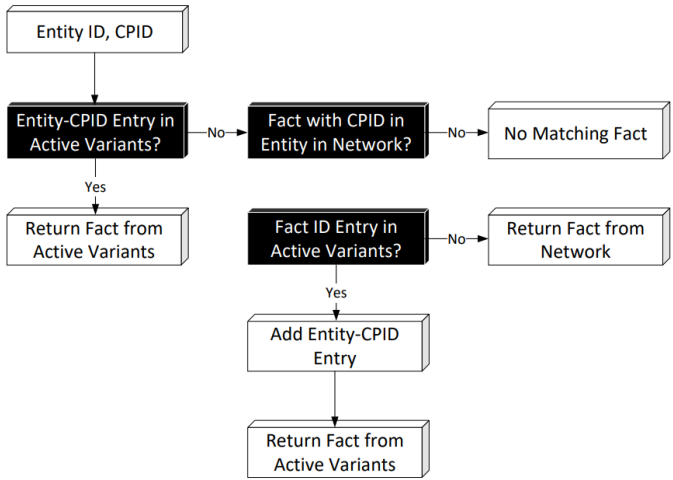

**Figure 20.** Retrieving a fact by the ID of a containing entity and the fact's CPID.

The pre- and post-ignore parameters, which provide the ability for a rule to ignore required pre- and postconditions, if they are not found, were also modified. The options are dependent on what is to be skipped. Facts may be skipped when applying postconditions, such that the specified facts' values are never altered. Common property identifiers can also be specified, and all facts with the specified common property identifier will not be altered.

Common property identifiers can also be set to be created if missing (referred to as 'post-create'). When retrieving a postcondition's target fact ID from a reality path, if a fact with the specified common property identifier (CPID) is not present in the specified location, and the CPID is in the post-create list, then a fact is created with that CPID. This process is shown in Figure 21. This fact will be placed only in the active variants list—not the base network—as it is associated with only the individual reality path.

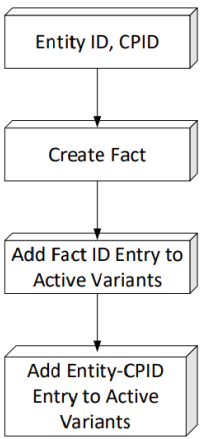

**Figure 21.** Adding a post-create fact.

Preconditions can also be configured to be skipped in the scenario parameters. By default, when a reality path does not have a fact required by a precondition, the corresponding rule is immediately disregarded, as its requirements cannot be satisfied. If the precondition is set to be skipped, however, and the required fact cannot be found, then the algorithm immediately stops evaluating that precondition and continues with the next precondition in the rule. This allows granularity when testing and during rule creation, as certain preconditions can be configured to be evaluated only if their requirements are present, rather than requiring the user to create multiple rules that have the same postconditions and manage them within the rule selection menu.

The active variant facts system is an area of potential improvement in the future. The current system is not as efficient as it could be, in terms of memory usage, processing, and I/O operations. When active variant facts' values are altered, they remain in the active variants collection within the reality path, even if the new value matches the corresponding fact's value in the base network. Thus, the active variant facts collection of every reality path only grows in size, and possibly holds facts with the same ID and value as facts in the base network. This may have history preservation benefits, but it increases the memory required by SONARR.

In order to facilitate accessing active variants by either their fact ID or by the IDs of their containing entity and CPID, both types of reference keys must be created to point to the same active variant fact object. In order to associate an entity ID and CPID to an active variant fact ID, the base entity is retrieved from the base network and its facts are iterated through, looking for the matching CPID. Since facts are referenced only by ID, when applying postcondition alterations, and an active variant fact is created upon the fact's first alteration, active variant facts are initially only referenced by the fact ID. The entity-common property entry is only added later, when the active variant is referenced by an entity-common property key. There is a processing impact, due to this, because of a subsequently discussed comorbid issue. Most of the entity-CPID entries are lost during serialization, resulting in this referencing work needing to be done many times.

I/O utilization is also not currently optimized. During development, the entity-common property entries for active variant facts were included during reality path serialization. This led to I/O bottlenecking, due to the tremendous amount of serialization and deserialization done by SONARR during processing. To mitigate this, only entity-common property entries pertaining to post-create facts are serialized, as they cannot be found in the base network. The way active variant facts are stored and retrieved is an area of potential future development, to improve RAM, CPU, and I/O efficiency.

The genericism aspect of fact values and pre-/postconditions would also benefit from further future expansion. Fact types could be added that use values of new object types. New object types would have their own requirements for comparison. For example, an object type representing the version of an operating system could be compared with less than and greater than operators. Given that a user may want to represent these values as strings but compare them in use case specific ways, conditions could be expanded on by allowing comparators to be overridden during creation.

3.3.1.2.2. Algorithm Enhancements

When the SONARR algorithm evaluates a connection within a reality path, it iterates through the rules in the order specified in the scenario parameters. It continues to the next reality path connection evaluation immediately upon a traversal rule triggering, or it halts the reality path completely, upon the failure to satisfy any rule evaluation. Additionally, evaluation is halted if the number of non-traversal rules triggered in the connection exceeds the maximum set in the scenario parameters. This means that no traversal rules were triggered and the reality path is stalled. This rule evaluation process has been changed to accommodate the generic logic.

Rule precondition assessment involves iterating through the rule's preconditions. For the rule to trigger, all preconditions must be satisfied. To satisfy a precondition, every requirement must match a fact, and the precondition's expression must evaluate to true. However, a precondition may be skipped if SONARR fails to find a required fact and this fact is in the scenario parameters' precondition ignore-if-missing list.

Rule postcondition assessment involves three parts, which are now described.

First, the algorithm gets the target facts from the reality path. Postconditions can either have one target fact—specified by a fact ID or by an entity ID and CPID—or several target facts, in which case all facts with the specified CPID are targeted. If a target fact is not found in an entity, but it is present in the scenario parameters' CPID post-create list, then the fact is created and added to the reality path's active variants list. If no target facts are found, the postcondition is skipped.

Second, the algorithm gets the fact values required for the postcondition expression. These facts must be present in the reality path. To ensure this, rule preconditions can be used to check for these facts. A precondition that exists to verify the presence of a postcondition should not be put into the precondition ignore-if-missing list.

Finally, the postcondition expression is evaluated, and the resulting value is set for the target facts. The target facts collected earlier are iterated through. A fact is skipped if it is in the fact post-ignore list or if the fact's CPID is in the CPID post-ignore list. Otherwise, the fact ID is used to create a variant fact within the reality path. This variant fact receives the postcondition's result value.

After all of the postconditions for a rule are applied, the entities in the current connection are updated with the active variants from the reality path. While the active variants are used by the algorithm, the connections need to store a snapshot of the active variants for later visualization.

## 4. Experimentation

The experiments in this section utilize network models to demonstrate and evaluate the expanded generic logic assessment functionality. Performance tests were also conducted to compare previous SONARR versions' multi-threaded, single-compute Boolean logic with the new multi-threaded, multi-compute generic logic. Those tests are also presented in this section.

### 4.1. Generic Logic Experimentation

Two networks were used to illustrate a vulnerability labeling use case. The first example, model 1, is a relatively simple network that starts at a container representing a common room in which four computers are stationed. Each of these computers has two facts that represent the name and version of a software instance on it. The rules assess the values of the facts within these computers against predefined vulnerability requirements and label them with corresponding fictitious CVEs. This labeling is performed by creating a fact in the vulnerable container with a common property identifier representing the CVE. A depiction of model 1 is shown in Figure 22.

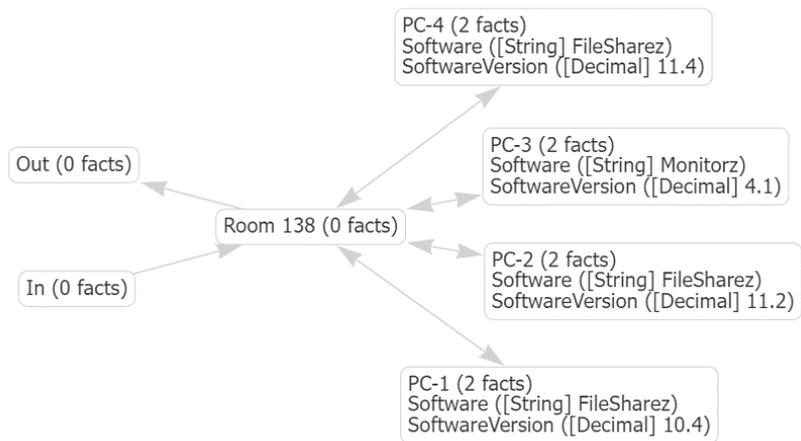

**Figure 22.** Depiction of model 1.

The second example, model 2, has a configuration of four computers connected to a database, with one of the computers being an administrative terminal for the database. Three real CVEs were simplified and implemented in this model's ruleset. A depiction of model 2 is shown in Figure 23.

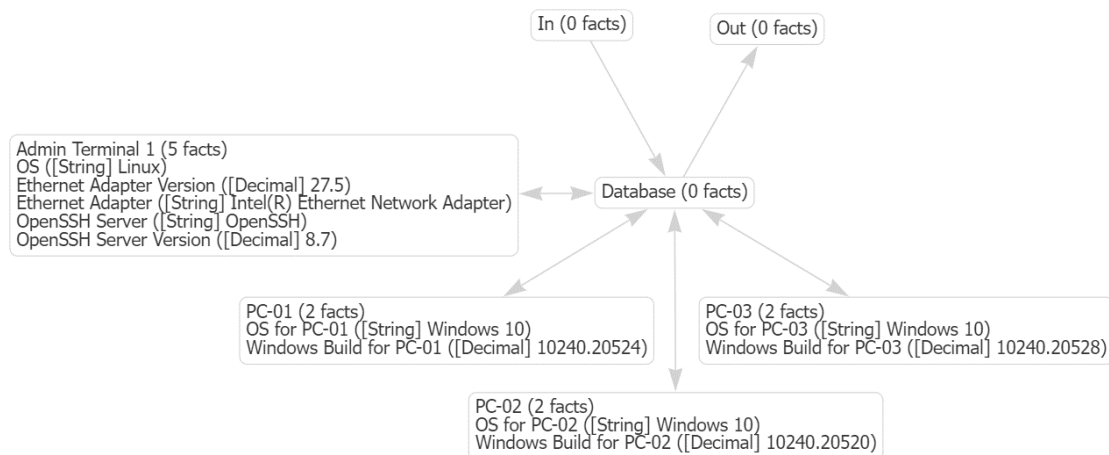

**Figure 23.** Depiction of model 2.

These networks demonstrate the flexibility of the SONARR network objects and the new logic. The underlying objects are generic and allow a user to pursue many avenues when modeling a network for analysis. The detail level of the network is up to the designer.

The initial development and testing of this technology showed that the modeling process can be difficult for humans to accomplish effectively. This indicates that making improvements to the modeling system could be an important area for future work.

### 4.2. Multi-Compute Experimentation

To test the implementation of the updated multi-compute algorithm, three scenarios for a large network from [3] were run using the algorithm, with and without using the multi-compute functionality, to compare the processing performance of the old Boolean logic and the new generic logic. This network will be referred to as model 3 and is shown in Figure 24.

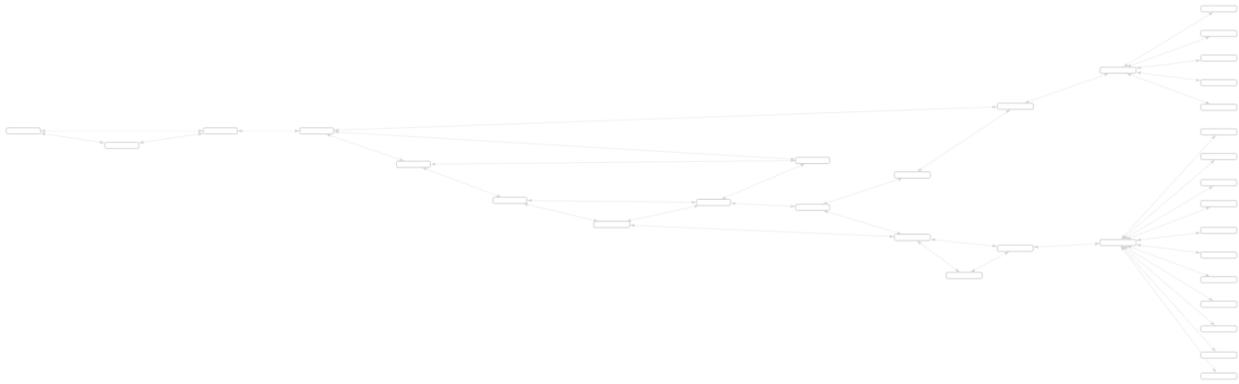

**Figure 24.** Depiction of model 3.

The model 3 network was designed for a performance test in [3]. It has only two rules: one to cross a link normally, and one to cross a link while closing it. Each link in the network has a fact indicating its open or closed status, which is changed to false when the second rule traverses it. Changing link facts to false, before running a scenario, allows SONARR to process specific sections of the network. Different sections of the network contain different connection architectures, allowing for varying traversal patterns, behaviors, and performances. Table 1 describes the three scenarios from [3] that were used to test the new logic and multi-compute.

**Table 1.** Previous scenarios for model 3 to test in the new logic and multi-compute.

|  | Scenario 1 | Scenario 2 | Scenario 3 |
| --- | --- | --- | --- |
| Start Container | Container 22 | Container 1 | Container 1 |
| End Container | Container 29 | Container 11 | Container 15 |
| Configuration | F57: F | F30, F33, F38: F | -- |
| Description | "big umbrella" | "middle section" | "whole network" |

Figures 25 to 27 illustrate the sections scenario 1, 2, and 3 processed, respectively.

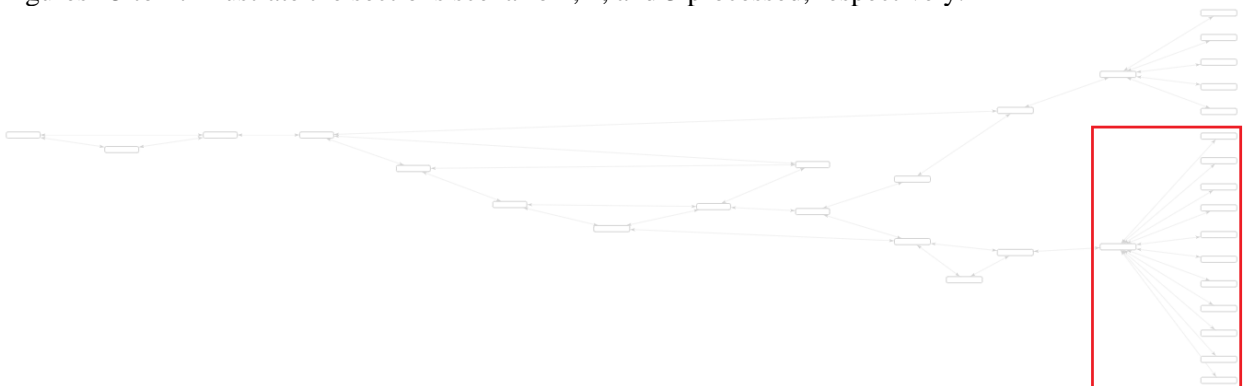

**Figure 25.** The 'big umbrella' processed in scenario 1.

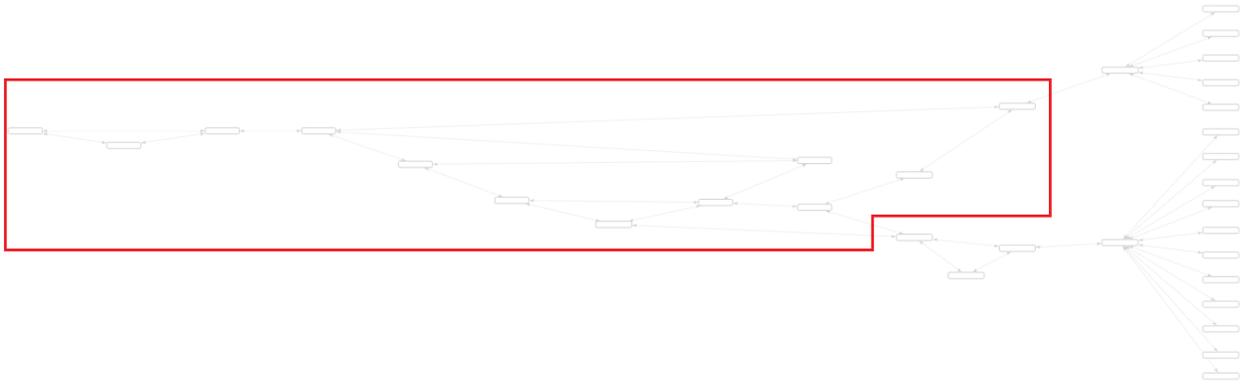

**Figure 26.** The 'middle section' processed in scenario 2.

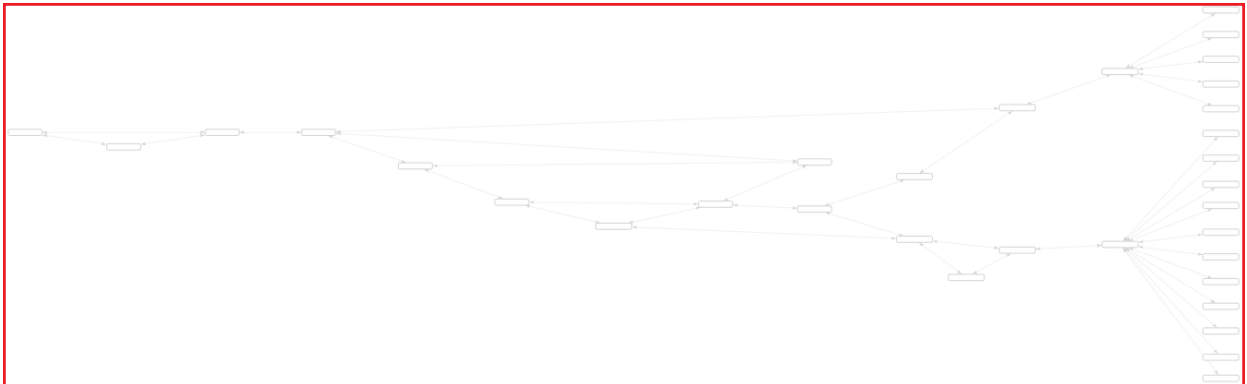

**Figure 27.** The 'whole network' processed in scenario 3.

## 5. Results & Discussion

Models 1 and 2 were run in SONARR from their start container to their end container. The models were designed with specific results anticipated and their correctness was manually validated.

Running model 1 from container 'In' to container 'Out' produces 65 final paths. The longest paths–with 10 connections and 12 triggered rules–are the important ones for this use case. These paths show the algorithm traversing from the 'In' container, through Room 138 to each PC, and then back out to Room 138, before reaching the 'Out' container.

During the visits to each PC, a rule assesses the two facts inside the entity: the software and the SoftwareVersion. For the CVE rule to trigger, the entity must contain a software fact with a string value of "FileSharez" and a SoftwareVersion fact with a decimal value of 11.2 or less. If those conditions are met, the postcondition runs and sets the CVE-2024-28394 common property identifier within the entity to true. It is important to note that, if there is not a fact with this common property identifier present, then nothing happens. Therefore, at the beginning of the SONARR run, a setting is configured to tell the algorithm to create a fact, if the CVE common property is not present in the postcondition's destination.

For model 1, only two PCs, out of the four, contain a CVE fact: PC-1 and PC-2. All the relevant paths met this criterion. Figures 28 to 31 illustrate the route taken by one of these paths. Additionally, Figures 31 and 30 show PC-1 and PC-2, respectively, with the expected CVE fact. Figures 29 and 28 show PC-3 and PC-4, respectively, without the CVE fact.

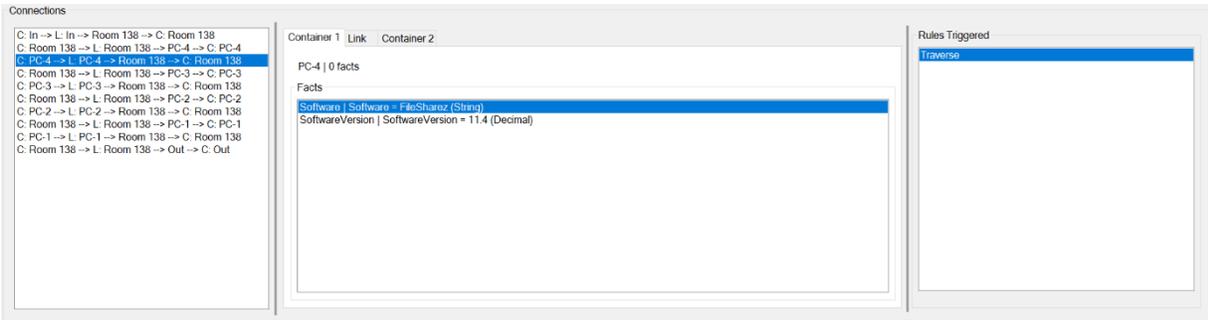

**Figure 28.** PC-4 container without the CVE fact in the relevant path from the model 1 results.

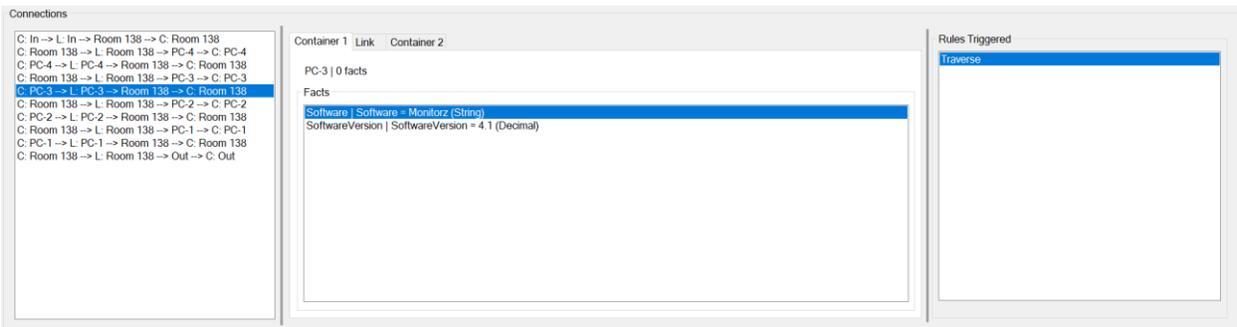

**Figure 29.** PC-3 without the CVE fact in the relevant path from the model 1 results.

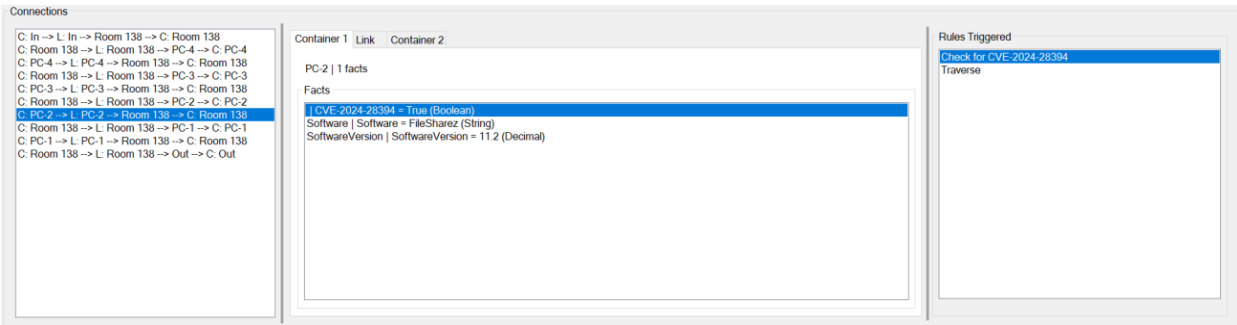

**Figure 30.** PC-2 container with the CVE fact in the relevant path from the model 1 results.

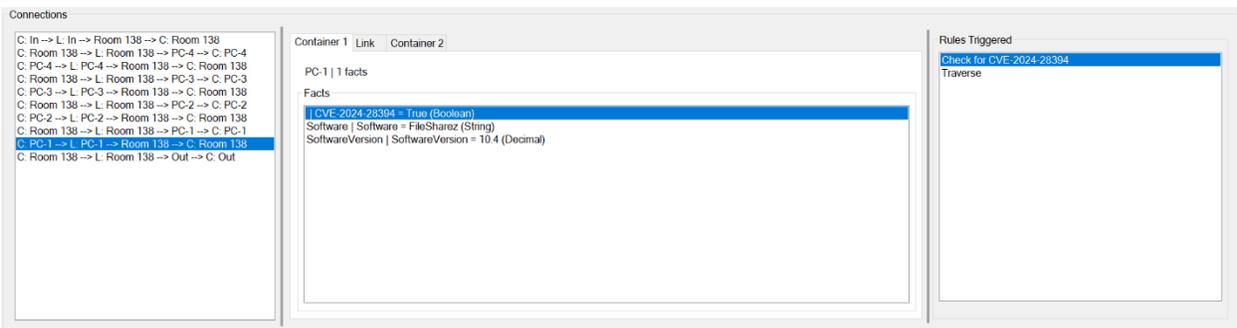

**Figure 31.** PC-1 container with the CVE fact in the relevant path from the model 1 results.

Model 2 has the same network configuration as model 1, so running model 2 from the 'In' container to the 'Out' container produces 65 final paths. Like model 1, the important paths for this case are again the longest paths, with 10 connections and 14 triggered rules. CVE-2024-43491, which identifies Windows 10 builds before 10240.20526 as vulnerable, is applicable to two PCs: PC-01 and PC-02. Additionally, there are two CVEs applicable to the admin terminal 1 entity. The first, CVE-2024-6409, identifies a vulnerability within an OpenSSH server, if it's version is between 8.7 and 8.8. The second, CVE-2024-218007, identifies a vulnerability of the Linux operating system for Intel(R) Ethernet Network Adaptors before version 28.3.

Manually analyzing the paths shows that SONARR successfully marked the vulnerable entities. Figures 32 to 35 illustrate the route SONARR took for one of the paths and the CVEs established for the vulnerable entities. Figure 32 shows that PC-03, which was not vulnerable to CVE-2024-43491, was not labeled with the vulnerability fact. Figure 34 shows the vulnerable PC-01 with its CVE fact, and Figure 33 shows the same for PC-02. Figure 35 shows the admin terminal having two CVE facts, one for the Intel network adaptor on the Linux operating system and the other for the vulnerable OpenSSH server. Table 2 provides the results for the two tests, including the general parameters set for each run, the completion time, and the final paths found by SONARR.

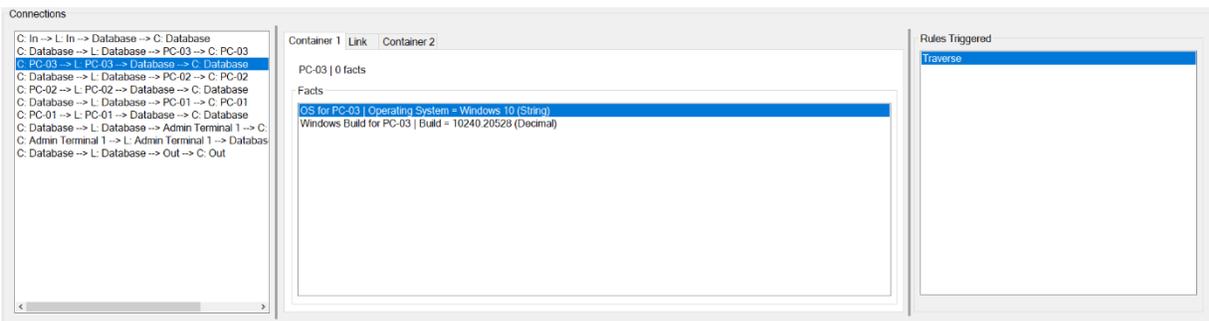

**Figure 32.** PC-03 container without a CVE fact in the relevant path from the model 2 results.

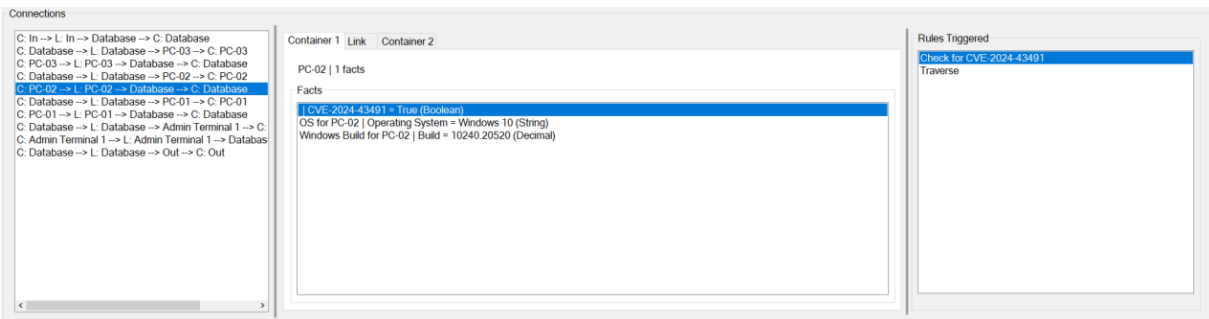

**Figure 33.** PC-02 container with the correct CVE fact in the relevant path from the model 2 results.

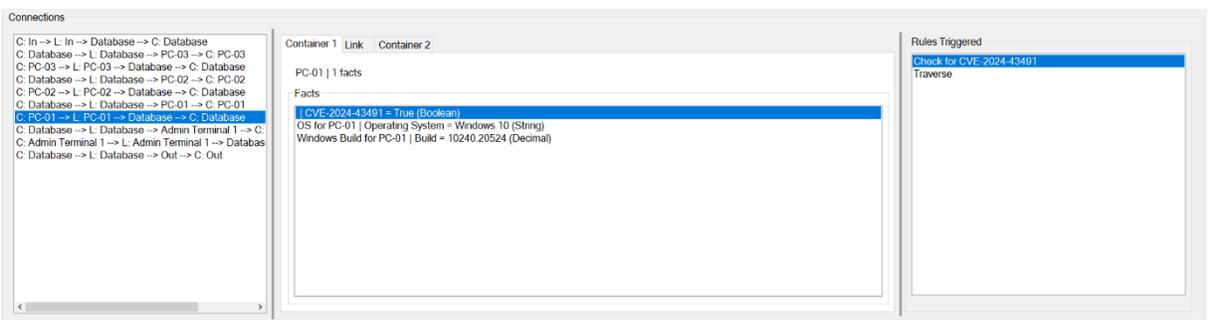

**Figure 34.** PC-01 container with the correct CVE fact in the relevant path from the model 2 results.

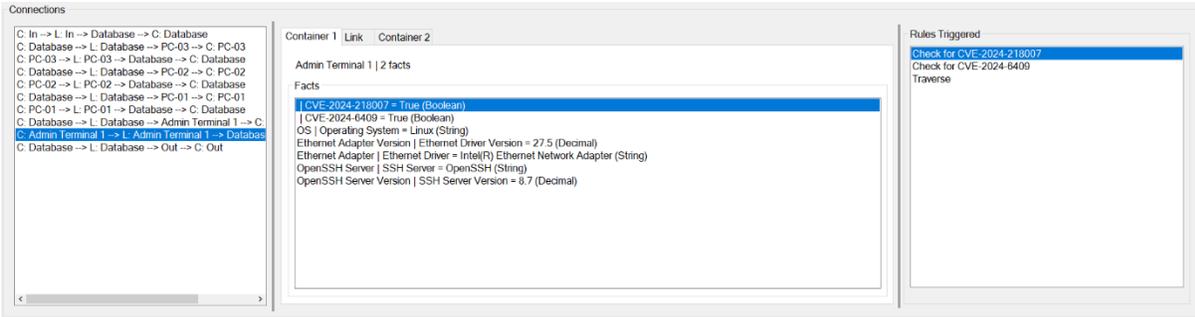

**Figure 36.** Admin Terminal container with the two correct CVE facts in the relevant path from the model 2 results.

**Table 2.** Processing results for the vulnerability labeling examples.

| Test | Model | Parameters | Time | Final Paths |
|---|---|---|---|---|
| 1 | 1 | Create fact(s) on postcondition<br>- CVE-2024-28394 | 00:00:00:01 | 65 |
| 2 | 2 | Create facts(s) on postcondition<br>- CVE-2024-43491<br>- CVE-2024-6409<br>- CVE-2024-218007 | 00:00:00:01 | 65 |

The point of these example network models is to demonstrate that, depending on how the network is configured, rules can be created using the implemented generic logic to match the prerequisites of CVEs in a more human readable way, as compared to the Boolean-only logic previously used. These CVE rules can then be used with a network model that consists of fact values of any kind (e.g., strings, decimals, Boolean, integers) to identify potential vulnerabilities within the system that the model represents. This is important as, the more human usable the framework is, the more intuitive and expandable the models are.

*5.1. Multi-Compute Results*

This section presents the results of the multi-compute testing. The multi-compute tests all use four nodes: one controller and three workers. The test results for scenario 1 are presented in Table 3. The test results for scenario 2 are presented in Table 4. Finally, processing times for Boolean logic using the single-compute algorithm (from [3]) and generic logic using the multi-compute algorithm are presented in Table 5. The results for scenario 3 are discussed subsequently.

**Table 3.** Processing results for ten scenario 1 ("big umbrella") runs. See Table 1.

| Test | Processing Time | Controller Results | Worker 1 Results | Worker 2 Results | Worker 3 Results | TOTAL |
|---|---|---|---|---|---|---|
| 1 | 24m 51s | 3,355,322 | 2,145,351 | 2,239,083 | 2,124,345 | 9,864,101 |
| 2 | 24m 55s | 3,392,609 | 2,299,462 | 2,148,244 | 2,023,786 | 9,864,101 |
| 3 | 27m 23s | 3,482,867 | 2,326,982 | 2,087,672 | 1,966,580 | 9,864,101 |
| 4 | 24m 58s | 3,359,212 | 2,246,733 | 2,312,999 | 1,945,157 | 9,864,101 |
| 5 | 24m 52s | 3,345,363 | 2,313,929 | 2,159,558 | 2,045,251 | 9,864,101 |
| 6 | 25m 2s | 3,273,588 | 2,132,786 | 2,158,043 | 2,299,684 | 9,864,101 |
| 7 | 27m 17s | 3,689,988 | 1,962,789 | 2,050,266 | 2,161,058 | 9,864,101 |
| 8 | 24m 54s | 3,306,904 | 2,209,095 | 2,163,881 | 2,184,221 | 9,864,101 |
| 9 | 28m 1s | 3,497,338 | 2,186,681 | 2,073,331 | 2,106,751 | 9,864,101 |
| 10 | 27m 3s | 3,674,387 | 2,027,269 | 2,065,346 | 2,097,099 | 9,864,101 |

| AVG | ~25m 55s | 3,437,758 | 2,185,108 | 2,145,842 | 2,095,393 | 9,864,101 |

Scenarios 1 and 2 had a natural limit on the number of possible paths they could find: 9,864,101 paths and 185,913 paths, respectively. Therefore, the processing time for these scenarios is the metric to focus on. The averages in Table 3 show that in scenario 1 final paths were found across the worker nodes somewhat evenly, and the workers managed to find all 9,864,101 paths in just under 26 minutes. The averages in Table 4 show a similar result for scenario 2, where the system found all 185,913 paths in just over a minute.

**Table 4.** Processing results for ten Scenario 2 ("middle section") runs. See Table 1.

| Test | Processing Time | Controller Results | Worker 1 Results | Worker 2 Results | Worker 3 Results | TOTAL |
|---|---|---|---|---|---|---|
| 1 | 1m 4s | 106,156 | 33,208 | 16,646 | 29,903 | 185,913 |
| 2 | 1m 4s | 84,315 | 59,985 | 1,640 | 39,973 | 185,913 |
| 3 | 1m 3s | 86,337 | 57,963 | 1,650 | 39,963 | 185,913 |
| 4 | 1m 6s | 108,395 | 20,671 | 42,618 | 14,229 | 185,913 |
| 5 | 1m 2s | 85,774 | 3,793 | 56,119 | 40,227 | 185,913 |
| 6 | 1m 5s | 88,975 | 36,255 | 49,054 | 11,629 | 185,913 |
| 7 | 1m 2s | 104,278 | 3,311 | 40,077 | 38,247 | 185,913 |
| 8 | 1m 5s | 85,434 | 10,737 | 33,939 | 55,803 | 185,913 |
| 9 | 1m 4s | 87,650 | 49,316 | 8,874 | 40,073 | 185,913 |
| 10 | 1m 2s | 87,886 | 5,658 | 40,043 | 52,326 | 185,913 |
| AVG | ~1m 3s | 92,520 | 28,090 | 29,066 | 36,237 | 185,913 |

**Table 5.** Scenario 1 and 2 processing times for Boolean logic tests and new generic logic tests.

| Scenario | Final Paths | Single-Compute | Multi-Compute |
| | | Boolean Logic | Generic Logic |
|---|---|---|---|
| 1 | 9,864,101 | 9m 3s | 1h 3m 45s | 25m 55s |
| 2 | 185,913 | 18s | 1m 37s | 1m 3s |

The time comparisons in Table 5 show that the Boolean logic is much faster, as expected. It also shows that there is a significant increase in processing time when using generic logic in single-compute, and there is a comparable decrease experienced when using multi-compute for the generic logic, both of which are also expected. While the Boolean logic is faster, it does not provide the expandability that the generic logic does. The multi-compute capabilities can bring the generic logic processing times to speeds similar to that of the Boolean logic, albeit requiring more computing resources to do so.

The scenario 3 performance tests are presented in Table 7. Only five tests were performed, as each one takes 24 hours to complete.

**Table 7.** Processing times for various Scenario 3 ("whole network") runs. See Table 1.

| Test | Processing Time | Controller Results | Worker 1 Results | Worker 2 Results | Worker 3 Results | TOTAL |
|---|---|---|---|---|---|---|
| 1 | 24h | 26,078,037 | 0 | 0 | 0 | 26,078,037 |
| 2 | 24h | 19,289,151 | 0 | 14,354,146 | 0 | 33,643,297 |
| 3 | 24h | 26,414,514 | 0 | 0 | 0 | 26,414,514 |
| 4 | 24h | 28,061,163 | 0 | 0 | 34,244,092 | 62,305,255 |
| 5 | 24h | 31,001,387 | 0 | 5 | 0 | 31,001,392 |
| AVG | 24h | 26,168,850 | 0 | 2,870,830 | 6,848,818 | 35,888,499 |

Due to the architectural nature of the model network used in this scenario, the initial paths provided to each worker node dictated whether it found final paths or not. In some cases, the workers receive work and process billions of potential paths, yet do not find a single path from C1 to C15. The multi-compute is doing its job, but the task is too large to complete. This is expected, as this scenario has no logic to cull potential paths during processing; it simply attempts to find every path from C1 to C15. This lack of found paths from workers is not seen in scenario 1 or 2, because the processed areas are significantly smaller in those scenarios.

With that said, over the course of five 24-hour runs, the controller was shown to find results in every test, averaging 26,168,850 final paths. Worker 1 only processed unfruitful paths, averaging zero final paths, worker 2 averaged 2,870,830 final paths, and worker 3 averaged 6,848,818. The overall average final paths found across these five tests is 35,888,499. These results are slightly greater than the results from the multi-threaded, single-compute Boolean logic presented in [3]. The increase in the number of paths processed behind-the-scenes means that, if run to completion, the multi-compute generic logic would finish earlier than the other algorithm for larger workloads. The results for these two algorithms is presented in Table 8.

**Table 8.** Scenario 3 results for Boolean logic test and new generic logic test.

| Scenario | Processing Time | Single-Compute Boolean Logic | Multi-Compute Generic Logic |
|---|---|---|---|
| 3 | 24h | 24,278,791 | 35,888,499 |

The average results that each node collected during the multi-compute runs for each of model 3's scenarios are presented in Table 9. These averages show the multi-compute implementation successfully distributed and processed work across the workers, which is important to remember when assessing the empty workers' results for the third scenario.

**Table 9.** Average results data for the tests on multi-compute performance scenarios.

| Scenario | Tests | Time | Controller Results | Worker 1 Results | Worker 2 Results | Worker 3 Results | Total |
|---|---|---|---|---|---|---|---|
| 1 | 10 | ~25m 55s | 3,437,758 | 2,185,108 | 2,145,842 | 2,095,393 | 9,864,101 |
| 2 | 10 | ~1m 3s | 92,520 | 28,090 | 29,066 | 36,237 | 185,913 |
| 3 | 5 | 24h | 26,168,850 | 0 | 2,870,830 | 6,848,818 | 35,888,499 |

## 6. Conclusions & Future Work

This paper has provided an overview and evaluation of the updates to the SONARR technology. A new generic logic capability was added to the rule-fact system, allowing facts in networks to be created with any value type (e.g., Boolean, string, integer, and decimal), making the model creation and result analysis more human understandable and intuitive. The presented examples and experimentation demonstrate the possible applications and expandability achievable via this method of system modeling.

Second, the algorithm's performance was improved through the implementation of multi-compute functionality. The multi-compute algorithm was shown to decrease processing time for large workloads, because of an increase in available processing resources. Notably, it increases the processing time for small workloads, which is not surprising as the multi-compute implementation adds computational overhead.

Areas for potential future work include exploring connection bypassing for logically linked entities, so that rules can assess two containers without having to unnecessarily incorporate relay devices between them. Future work could also explore the benefits of using AI to make informed, trajectory-influencing decisions during processing, and integrate STIG configurations to facilitate network creation and verification.

**Acknowledgements**

This work utilized and built on the SONARR system, the development of which was funded by the U.S. Missile Defense Agency (contract # HQ0860-22-C-6003).

**Appendix A. Enlarged Diagrams**

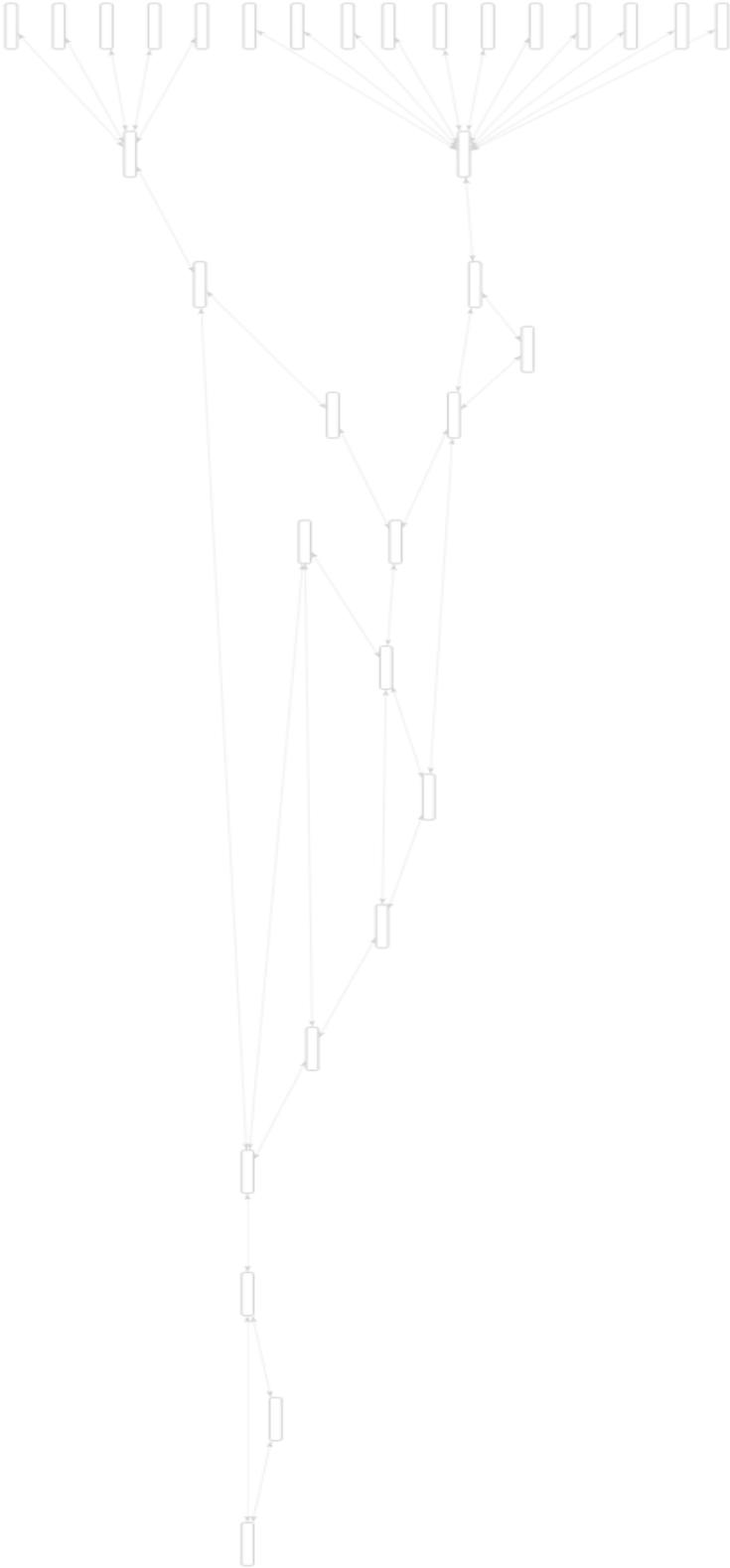

**Figure A1.** Depiction of model 3 (enlargement of Figure 24).

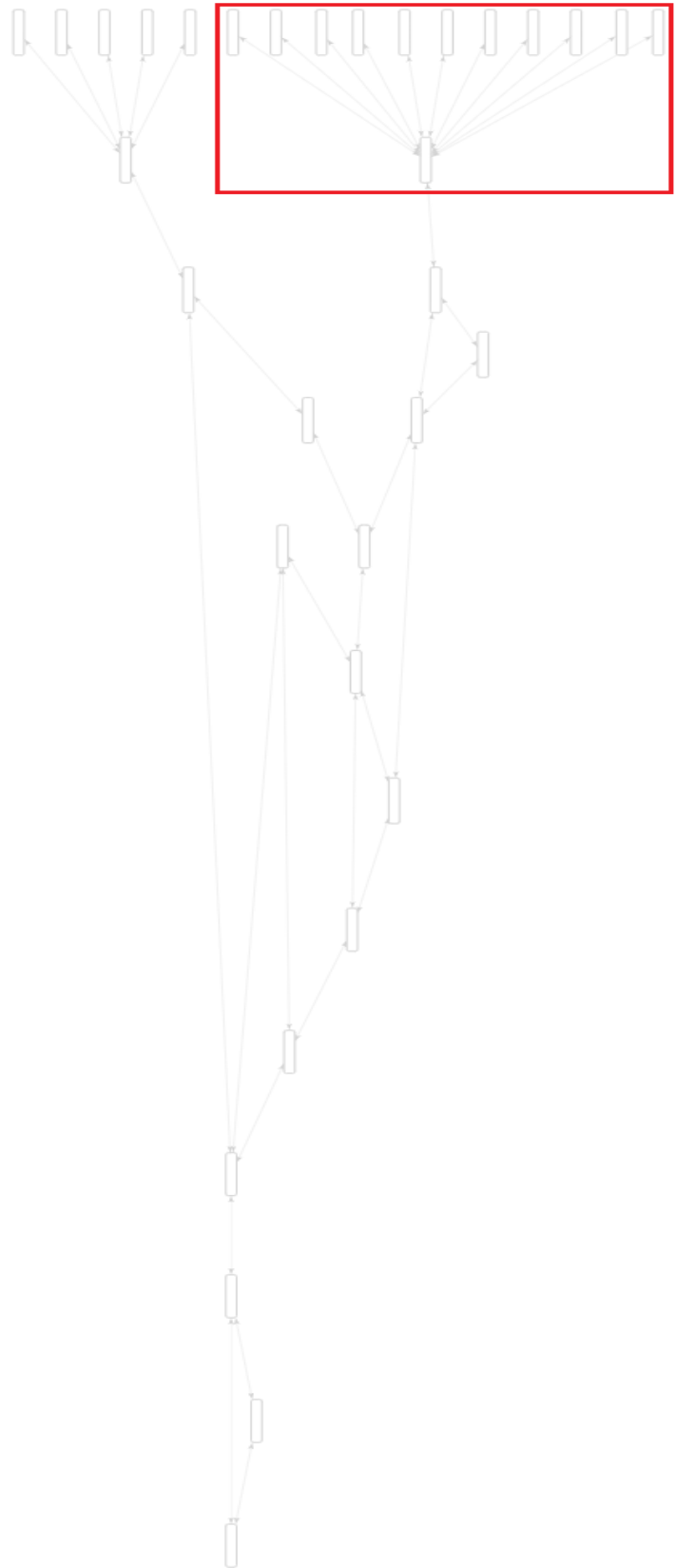

**Figure A2.** The "big umbrella" processed in scenario 1 (enlargement of Figure 25).

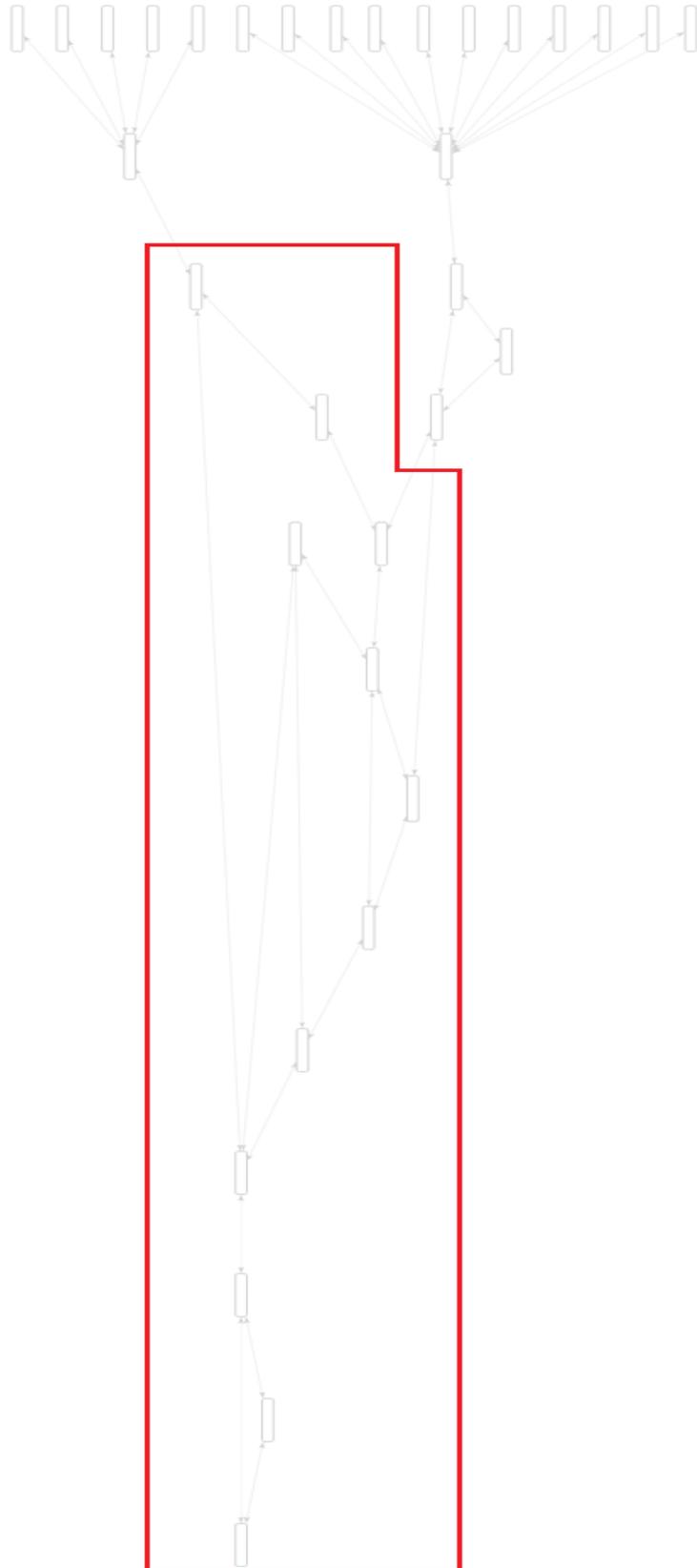

**Figure A3.** The "middle section" processed in scenario 2 (enlargement of Figure 26).

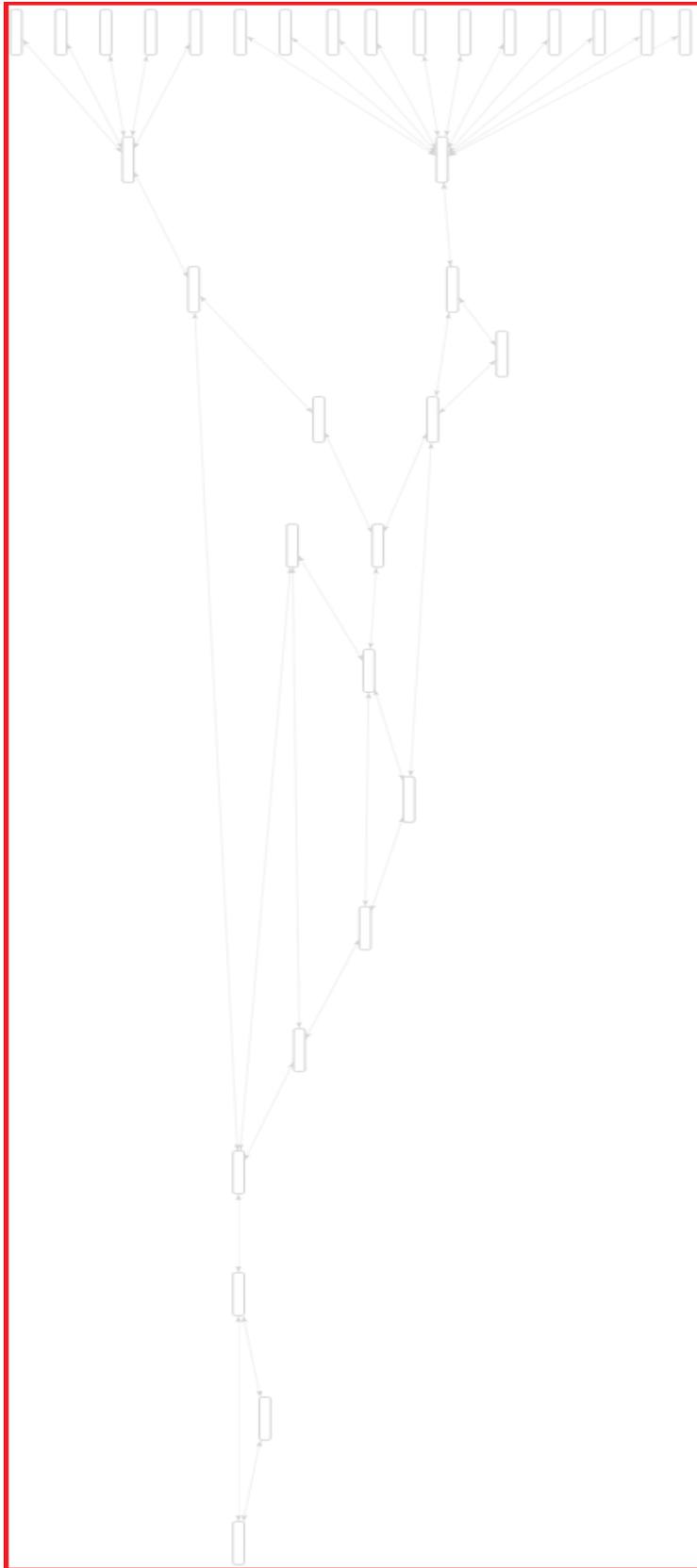

**Figure A4.** The "whole network" processed in scenario 3 (enlargement of Figure 27).